\documentclass[12pt]{iopart}
\usepackage{iopams}
\usepackage{amssymb}
\usepackage{esint}
\usepackage{graphicx}
\usepackage{bbold}
\usepackage[normalem]{ulem}
\usepackage{comment}
\usepackage{cite}

\usepackage{setspace}

\usepackage{breakurl}
\usepackage{seqsplit}
\usepackage[hidelinks,colorlinks]{hyperref}
\hypersetup{citecolor=[rgb]{0.25,0.14,0.63}}
\hypersetup{urlcolor=[rgb]{0.25,0.14,0.63}}
\hypersetup{linkcolor=black}

\makeatletter

\usepackage{color}

\begin{document}
\title[Finite-time quantum Otto engine]{A finite-time quantum Otto engine with tunnel coupled one-dimensional Bose gases}

\author{V.V. Nautiyal, R. S. Watson, 
and K. V. Kheruntsyan}

\address{School of Mathematics and Physics, University of Queensland, Brisbane, Queensland 4072, Australia}


\begin{abstract}
 We undertake a theoretical study of a finite-time quantum Otto engine cycle driven by inter-particle interactions in a weakly interacting one-dimensional Bose gas in the quasicondensate regime. Utilizing a $c$-field approach, we simulate the entire Otto cycle, i.e. the two work strokes and the two equilibration strokes. More specifically, the interaction-induced work strokes are modelled by treating the working fluid as an isolated quantum many-body system undergoing unitary evolution. The equilibration strokes, on the other hand, are modelled by treating the working fluid as an open quantum system tunnel-coupled to another quasicondensate which acts as either the hot or cold reservoir, albeit of finite size. We find that, unlike a uniform 1D Bose gas, a harmonically trapped quasicondensate cannot operate purely as a \emph{heat} engine; instead, the engine operation is enabled by additional \emph{chemical} work performed on the working fluid, facilitated by the inflow of particles from the hot reservoir. The microscopic treatment of dynamics during equilibration strokes enables us to evaluate the characteristic operational time scales of this Otto chemical engine, crucial for characterizing its power output, without any \emph{ad hoc} assumptions about typical thermalization timescales. We analyse the performance and quantify the figures of merit of the proposed Otto chemical engine, finding that it offers a favourable trade-off between efficiency and power output, particularly when the interaction-induced work strokes are implemented via a sudden quench. We further demonstrate that in the sudden quench regime, the engine operates with an efficiency close to the near-adiabatic (near maximum efficiency) limit, while concurrently achieving maximum power output.

\end{abstract}
\maketitle

\section{Introduction}\label{sec:Introduction}

Quantum heat engines (QHEs) \cite{koch2022making, rossnagel2016single, simmons2023thermodynamic, bouton2021quantum, klatzow2019experimental, fogarty2020many, keller2020feshbach, li2018efficient, singh2020optimal} provide a concrete platform for understanding the fundamental laws of thermodynamics in the quantum domain \cite{brandao2015second, masanes2017general}. Their exploration has recently expanded to include interacting many-particle systems \cite{mukherjee2021many, halpern2019quantum, chen2019interaction, watson2024interaction, herrera2023correlation, latune2020collective}, hence offering access to quantum many-body effects such as entanglement, quantum coherence, and correlations. Such quantum effects can be exploited for demonstrating either a form of quantum advantage \cite{jaramillo2016quantum} or a uniquely quantum functionality of QHEs not accessible classically \cite{koch2022making,bouton2021quantum}.

In QHEs that rely on interacting many-body systems as their working fluid, the strength of interparticle interactions provides a tool for engine operation not available in noninteracting systems.
For example, it is possible to tune or quench the interaction strength to either do work on, or extract work from, the working fluid, analogous to volumetric compression and expansion work strokes in a conventional Otto engine cycle \cite{chen2019interaction, watson2024interaction, keller2020feshbach,li2018efficient}. Alternatively, one can quench the interaction strength to change the internal energy of the system without exchanging heat through coupling it to a thermal reservoir in a conventional Otto engine. This energy can then be utilized to extract work from the working fluid, as was recently demonstrated in Refs. \cite{simmons2023thermodynamic, koch2022making}.

 Recent studies have also focused on optimising the finite time performance of interaction-driven many-body engines \cite{li2018efficient, keller2020feshbach, fogarty2020many, williamson2023many, solfanelli2023quantum }. 
In general, the maximum efficiency for any QHE is attained when the work strokes are executed in the near-adiabatic or quasistatic limit, in accordance with the quantum-adiabatic theorem \cite{abah2019shortcut, hartmann2020many}. This leads to zero power output due to infinitely long engine driving times \cite{li2018efficient, abah2019shortcut, mukherjee2021many}. On the other hand, if the work strokes are rapid, the production of irreversible work due to non-adiabatic excitations results in a significant reduction of efficiency \cite{keller2020feshbach,shiraishi2016universal,abah2019shortcut}. Therefore, a major challenge in quantum thermodynamics is to design QHEs that provide a favourable trade-off between efficiency and power output, meaning they can operate with maximum or near-adiabatic efficiencies while providing non-zero power output in finite time \cite{campbell2017trade, campisi2016power, shiraishi2016universal}.

Recent developments aimed at optimising efficiency in finite-time operations have predominantly focused on employing shortcut to adiabaticity (STA) protocols \cite{guery2019shortcuts,abah2019shortcut, keller2020feshbach}. However, given that STA protocols come with certain drawbacks such as modulation instability \cite{keller2020feshbach, li2018efficient}, additional energetic costs for implementation \cite{calzetta2018not}, and challenges in experimental realization \cite{chen2010transient}, the exploration of alternative and preferably simpler approaches to operate at near-maximum efficiency while maintaining non-zero power output becomes important. Furthermore, when evaluating the power, the dynamics of equilibration of the working fluid with the reservoir during the thermalization strokes are often ignored. This is reasonable when assuming that the reservoir size is infinite and that the thermalization time with such a reservoir is much shorter than the duration of the work strokes
\cite{keller2020feshbach, li2018efficient, boubakour2023interaction, li2021shortcut, ccakmak2019spin, rezek2006irreversible, kosloff2017quantum}. However, these assumptions may not hold true in laboratory experiments, where the system typically interacts with a finite-sized reservoir, meaning that the characteristic thermalization time of the system with the reservoir becomes important for working out the power output of an engine.

In this work, we conduct a theoretical investigation of an experimentally realizable quantum Otto engine driven by the quench of inter-particle interactions in a weakly interacting Bose gas in the quasicondensate regime. In addition to conducting microscopic simulations of the interaction-driven work strokes of such an engine, we simulate the equilibration strokes of the working fluid with the reservoir. This is done by treating the working fluid as an open quantum system in thermal and diffusive contact with another, larger quasicondensate serving as the reservoir. Through this microscopic treatment of the equilibration strokes, we evaluate characteristic operational timescales for these strokes which enables us to quantify the power output of the engine for experimentally realistic time scales of the full Otto cycle. 
From this analysis, we demonstrate that, unlike in the uniform 1D Bose gas investigated in Refs.~\cite{chen2019interaction, watson2024interaction}, operation as a \textit{heat} engine is not possible for the harmonically trapped system. However, allowing for additional chemical work, in the form of particle flow from the hot reservoir to the working fluid, enables engine operation. In this scenario, the engine cycle may instead be considered as an effective \textit{chemical} engine cycle. Such a cycle is shown to possess a favourable trade-off between power output and efficiency when the work strokes are implemented via a sudden quench of the interaction strength. We show that within the sudden quench regime, the engine functions with efficiency nearing the limit of near-adiabatic (near-maximum) efficiency, while attaining maximum power output.

The article is organised as follows: In Section \ref{sec:2}, we introduce our model of the Otto cycle and describe the $c$-field approach that we use to simulate the complete finite-time Otto cycle. Following this, in Section \ref{sec:3}, we focus on the unitary work strokes and identify the timescales governing a sudden quench, an intermediate-time quench, and a near-adiabatic (quasistatic) quench. The knowledge of these timescales will enable us to evaluate the trade-off between efficiency and power when we analyse the finite-time performance of the engine. Next, in Section \ref{sec:4}, we shift our focus to the equilibration strokes with the reservoir. In this section, we investigate various dynamical scenarios governing the equilibration strokes and evaluate the operational timescales of these strokes, which will help us quantify the power output of the full engine cycle in an experimentally realisable scenario. In Section \ref{sec:5}, we utilize the findings of the previous sections to evaluate the performance of the proposed interaction-driven chemical Otto engine in its full four stroke cycle. We quantify figures of merit of the engine, such as the power output, efficiency, and the trade-off between power and efficiency, as we increase the duration of the work strokes from the sudden quench regime to the near-adiabatic (quasistatic) regime. Finally, in Section \ref{sec:conclusion}, we summarise our findings and discuss the outlook.

\section{The Otto cycle} \label{sec:2} 
\begin{figure}[t!]
  \centering
  \includegraphics[scale=0.7]{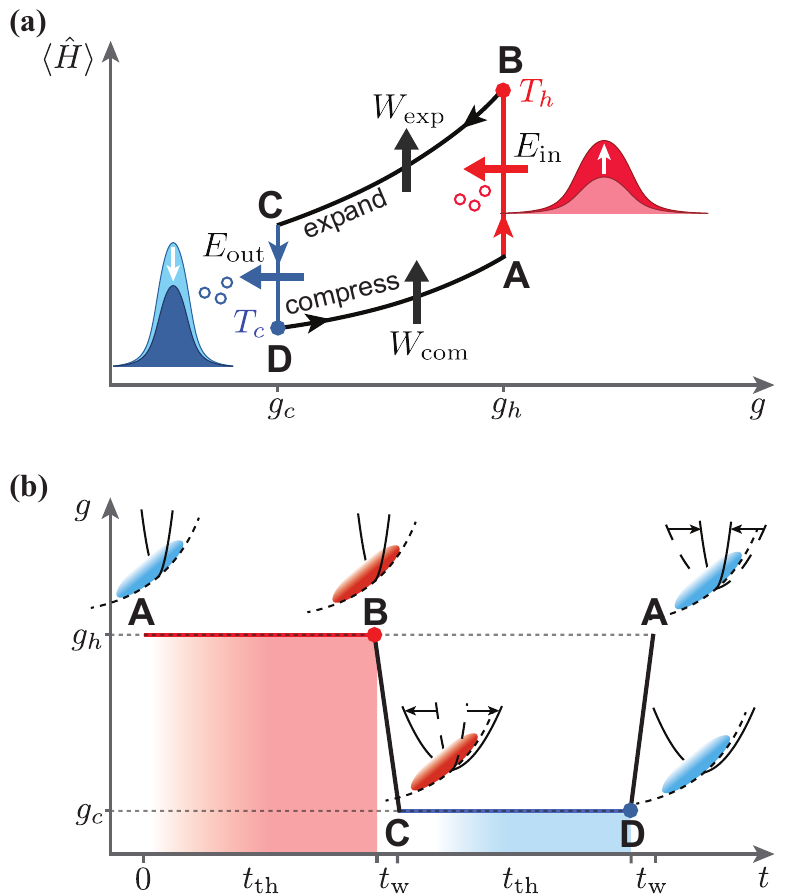}
\caption{
An interaction-driven quantum Otto cycle with a harmonically trapped 1D Bose gas as the working fluid. Panel (a) shows a schematic diagram of the four strokes of the Otto cycle in the working fluid energy $\langle \hat{H}\rangle$ versus interaction strength $g$ parameter space. For further details and notations, see text. Panel (b) illustrates the time sequence of the interaction strength quench (not to scale) during the four strokes of the Otto cycle: two thermalization strokes of duration $t_{\mathrm{th}}$, and two work strokes of duration $t_{\mathrm{w}}$. As $g\simeq 2\hbar \omega_{\perp}a$, the change in $g$ can be achieved via an increase or reduction of the frequency of transverse confinement $\omega_{\perp}$, which then makes the work strokes \textbf{DA} and \textbf{BC} analogous, respectively, to (transverse) compression or expansion strokes of the regular, volumetric Otto cycle. }

  \label{fig:Fig1}
\end{figure}

The Otto engine cycle, demonstrated schematically in Fig.~\ref{fig:Fig1}, is widely studied in the field of quantum thermodynamics due to its relative simplicity \cite{Myers2022quantum}, in addition to being the closest model to real-world engine cycles.
It consists of two unitary work strokes alternated with two isochoric equilibration strokes in which the working fluid is coupled to external thermal reservoirs. In particular, the unitary work strokes, denoted \textbf{BC} and \textbf{DA} in Fig.~\ref{fig:Fig1}(a), correspond to volumetric compression and expansion, respectively, and are implemented via an external control parameter over the working fluid. The equilibration strokes \textbf{AB} and \textbf{CD} consist of coupling the working fluid to hot and cold thermal reservoirs, at temperatures $T_h$ and $T_c$, respectively, while maintaining constant volume. In the following subsections, we will describe our model and the implementation of the individual strokes, as well as their combined operation in a full four-stroke cycle of the Otto engine.

\subsection{The Model}
In this work, we consider a working fluid consisting of a harmonically trapped ultracold 1D Bose gas in the weakly interacting quasicondensate regime \cite{Petrov_2000_Regimes, Mora-Castin-2003,kheruntsyan2005finite, kheruntsyan2003pair, garrett2013condensation, clade2009observation, Jacqmin_2011_subpoissonian}. The Hamiltonian of this system ($s$), in second-quantized form, is given by 
\begin{equation}\label{eq:hamiltonian}
     \hat{H}_s = \int dx \ \hat{\Psi}_s^{\dag} \Biggl[- \frac{\hbar^2}{2m} \frac{\partial^2}{\partial x^2} \ + \frac{1}{2} m \omega^2 x^2 + \frac{g_s}{2} \hat{\Psi}_s^{\dag} \hat{\Psi}_s    \Biggr] \hat{\Psi}_s,
\end{equation}
where $\omega$ is the longitudinal trapping frequency, $m$ is the particle mass, and $\hat{\Psi}_s^\dagger(x)$ and $\hat{\Psi}_s(x)$ are the bosonic field creation and annihilation operators, respectively. Further, $g_s$ is the strength of repulsive ($g_s>0$) interparticle interactions within the working fluid that can be related to the frequency of transverse confinement, $\omega_\perp$, and the three-dimensional $s$-wave scattering length, $a$, via the relationship $ g_s \!\simeq\! 2 \hbar \omega_{\perp}a$, valid away from confinement-induced resonances \cite{olshanii1998atomic}.

 Modelling the work strokes of the Otto cycle corresponds to simulating the unitary, real-time dynamics of the working fluid governed by the Hamiltonian~(\ref{eq:hamiltonian}) in response to a change of an external parameter. In this work, we study an interaction-driven Otto cycle, enacted through control over the interaction strength $g_s$.
 In practice, the interaction strength can be tuned either by changing the scattering length $a$ using magnetic Feshbach resonance \cite{chin2010feshbach}, or by varying the frequency of the transverse confinement $\omega_\mathrm{\perp}$ \cite{schemmer2018monitoring}, both methods leading to identical results reported here. Tuning the interaction strength by increasing or reducing $\omega_\perp$ can be regarded, respectively, as transverse compression or expansion of the working fluid \cite{watson2024interaction}, offering an analogy to the conventional volumetric Otto cycle even when the interaction strength is changed via a magnetic Feshbach resonance. During compression stroke \textbf{DA}, work, $W_\mathrm{com}>0$, is done on the working fluid by increasing the interaction strength, whereas in expansion stroke \textbf{BC}, the interaction strength is decreased allowing work, $W_\mathrm{exp}<0$, to be done by the working fluid. The net work extracted in one complete cycle is thus $W = W_\mathrm{{com}} + W_\mathrm{{exp}}$, which must be negative, $W<0$, for the cycle to operate as an engine.

To model the equilibration strokes, on the other hand, we consider a coupled system in which the working fluid, described by the Hamiltonian (\ref{eq:hamiltonian}), is coupled to another, larger 1D quasicondensate which serves as the reservoir. The coupling to the hot and cold reservoirs is alternated between the compression and expansion work strokes, but apart from that we assume that the hot and cold reservoirs are described identically, except for their respective thermal equilibrium temperatures. More specifically, for modelling the equilibration strokes, we employ the tunnel-coupled model of two quasicondensates \cite{bayocboc2022dynamics} described by the following Hamiltonian:
\begin{equation}
\label{eqn:tunnelcoupled}
    \hat{H}_\mathrm{coupled} = \hat{H}_{s} + \hat{H}_{h(c)}  - \hbar J \int dx \  [\hat{\Psi}^{\dag}_{s} \hat{\Psi}_{{h(c)}} + \hat{\Psi}^{\dag}_{h(c)} \hat{\Psi}_{s}].
\end{equation}
Here, the subscript $s$ denotes the system or the working fluid, whereas the subscripts $h$ and $c$ denote the hot and cold reservoirs, respectively. The Hamiltonians $\hat{H}_{h(c)}$ for the hot (cold) reservoirs have the same structure as Eq.~(\ref{eq:hamiltonian}), except that the field operators $\hat{\Psi}_s(x)$ are replaced by $\hat{\Psi}_{h(c)}(x)$. The parameter $J$ characterizes the tunnel-coupling strength between the working fluid and the reservoir; it can be determined precisely in experiments by probing their mutual two-point phase-correlation function using matter-wave interferometry \cite{betz2011two}.

\subsection{Numerical method}

To implement the four strokes of the quantum Otto cycle, we utilize the numerical $c$-field method, developed for studying the finite-temperature dynamics of Bose gases \cite{Blakie_cfield_2008, bayocboc2022dynamics, thomas2021thermalization, bayocboc2023frequency, simmons2020quantum, blakie2008dynamics}. 
This involves decomposing the quantum field operator $\hat{\Psi}_i(x,t)$ into two distinct regions: the classical, or $c$-field region, composed of highly occupied low-energy modes that can be described by a complex field amplitude $\psi^{(\textbf{C})}_{i}(x,t)$, and the thermal region containing sparsely occupied high-energy modes that act as an effective thermal bath for the $c$-field. We now detail the numerical implementation of the entire finite-time Otto cycle using the $c$-field method.

~

\noindent \textit{Step 1: Unitary compression work stroke,} \textbf{DA}: We prepare the initial finite-temperature thermal equilibrium state of the working fluid at point \textbf{D} of Fig.~\ref{fig:Fig1} using the stochastic projected Gross-Pitaevskii equation (SPGPE) \cite{gardiner2003stochastic,rooney2012stochastic},
\begin{equation}
\label{eq:SPGPE}
    d\psi^{(\textbf{C})}_{s}(x,t) = \mathcal{P} ^{(\textbf{C})}\Biggl\{- \frac{i}{\hbar} \mathcal{L}^{(\textbf{C})}_s \psi^{(\textbf{C})}_{s}(x,t) dt  +\frac{\Gamma}{k_{B}T_{s}}(\mu_{s} - \mathcal{L}^{(\textbf{C})}_s)\psi^{(\textbf{C})}_{s}(x,t) dt + dW_{\Gamma}(x,t) \Biggr\},
\end{equation}
where $\psi^{(\textbf{C})}_{s}(x,t)$ is the classical field of the working fluid or the system $(s)$. In Eq.~(\ref{eq:SPGPE}), the parameters $\mu_s$ and $T_s$ refer to the chemical potential and temperature of the thermal region and determine the total number of particles in the $c$-field region of the working fluid, $\psi^{(\textbf{C})}_{s}(x,t)$. The operator $\mathcal{P}^{(\textbf{C})}$ is a projector operator that sets up the boundary between the $c$-field and thermal region, defined by a cut-off energy, $\epsilon_\mathrm{cut}$. The Gross-Pitaevskii operator, $\mathcal{L}^{(\textbf{C})}_s$, is given by
\begin{equation}
\label{eqn:GPE_operator}
    \mathcal{L}^{(\textbf{C})}_s = - \frac{\hbar^2}{2m} \frac{\partial^2}{\partial x^2} + \frac{1}{2}m \omega^2 x^2 + g_s |\psi_s^{(\textbf{C})}(x,t)|^2,
\end{equation}
We note here that we have matched up the strength of interparticle interactions and the temperature of the working fluid to those of the cold reservoir, which is to be considered later (in stroke \textbf{CD}, see below), i.e., we have chosen $g_s = g_c$ and $T_s=T_c$ for the initial thermal equilibrium state of the working fluid.

Finally, the stochastic noise term, $dW_{\Gamma}(x,t)$, in Eq.~(\ref{eq:SPGPE}) corresponds to complex white noise, defined by the correlation \cite{blakie2008dynamics, gardiner2003stochastic, rooney2012stochastic},
\begin{equation}
\label{eqn:noise-correlation}
    \langle dW^*_{\Gamma}(x,t) dW_{\Gamma}(x',t) \rangle  = 2\Gamma \delta(x-x')dt.
\end{equation}
 where $\Gamma$ is the growth rate that characterises the strength of the coupling between the $c$-field and the effective thermal bath, with $\langle \cdot \rangle$ referring to stochastic averaging over a large number of independent stochastic realisations (trajectories). In practice, the growth rate $\Gamma$ may be chosen according to numerical convenience as it does not affect the final thermal equilibrium configuration \cite{blakie2008dynamics, rooney2012stochastic}.

At the end of this preparation stage, the working fluid is at point \textbf{D} in the Otto cycle diagram of Fig.~\ref{fig:Fig1}. To initiate the first (compression) work stroke of the Otto cycle, \textbf{DA}, we assume that the working fluid is isolated from all external reservoirs and its interaction strength is quenched from an initial value, $g_s=g_c$ to the final value $g_h$. We assume that the interaction quench is realised over a finite duration according to a linear protocol,
\begin{equation}\label{eq:g_com}
    g_s(t) = g_c+  \ (g_h-g_c) t / t_\mathrm{w},
\end{equation}
where $t_\mathrm{w}$ is the duration of the quench.
The working fluid now undergoes a unitary evolution, which is modelled by the following projected Gross-Pitaevskii equation (PGPE)
\cite{blakie2005projected,Blakie_cfield_2008}:
\begin{equation}
\label{eqn:PGPE}
    i \hbar \frac{\partial}{\partial t} \psi_\mathrm{s}^{(\textbf{C})}(x,t) = \mathcal{P} ^{(\textbf{C})} \Biggl\{ - \frac{\hbar^2}{2m} \frac{\partial^2}{\partial x^2} + \frac{1}{2} m \omega^2x^2 + g_s(t)|\psi_\mathrm{s}^{\textbf{(C)}}|^2 \Biggr\}.
\end{equation}
Through this stroke, mechanical work $W_\mathrm{com} >0$ is done on the working fluid. The quantity $W_\mathrm{com}$ is computed numerically by evaluating the change in the Hamiltonian energy of the working fluid after completion of the stroke at point \textbf{A}, i.e, $W_\mathrm{com}=\langle \hat{H}_s \rangle_{\textbf{A}}-\langle \hat{H}_s \rangle_{\textbf{D}}$.
The duration $t_\mathrm{w}$ in which the compression stroke \textbf{DA} is completed determines the state of the working fluid at the end of the stroke. If this compression stroke is executed via a sudden quench, the working fluid at point \textbf{A} will be in a highly non-equilibrium state with no definable temperature.
In contrast, if the compression stroke is carried out using a slow quasistatic quench, then the working fluid at point \textbf{A} will have a definite temperature, $T_s > T_c$.

~

\noindent\textit{Step 2: Equilibration with hot reservoir,} \textbf{AB}: Upon completion of the unitary compression stroke \textbf{DA} at time $t_{\mathrm{w}}$, we next model the subsequent equilibration stroke \textbf{AB} with the hot reservoir by switching on the tunnel coupling between the working fluid and the hot reservoir. In an experimental setup, the tunnel-coupled system of 1D quasicondensates, or the 1D bosonic Josephson junction, can be realized using a quantum degenerate Bose gas confined in a tunable double well potential in the transverse direction \cite{betz2011two}. The dynamics of the working fluid coupled to the hot reservoir are now modelled using the coupled PGPEs \cite{bayocboc2022dynamics},   
\begin{equation}
\label{eqn:coupled_PGPEs_a}
        i \hbar \frac{\partial}{\partial t} \psi^{(\mathbf{C})}_{\mathrm{s}}(x,t) = \mathcal{P}^{(\mathbf{C})} \Biggl\{ - \frac{\hbar^2}{2m} \frac{\partial^2}{\partial x^2} + \frac{1}{2}m \omega^2 x^2 +  g_s|\psi^{(\mathbf{C})}_{\mathrm{s}}|^2 - \hbar J \psi^{(\mathbf{C})}_{h}(x,t) \Biggr\}, 
         \end{equation}
 \begin{equation}       
    i \hbar \frac{\partial}{\partial t} \psi^{(\mathbf{C})}_{h}(x,t) = \mathcal{P}^{(\mathbf{C})}  \Biggl\{ - \frac{\hbar^2}{2m} \frac{\partial^2}{\partial x^2} + V_h(x)   + g_h|\psi^{(\mathbf{C})}_{h}|^2 - \hbar J \psi^{(\mathbf{C})}_{\mathrm{s}}(x,t) \Biggr\},
    \label{eqn:coupled_PGPEs_b}
    \end{equation}
where, $\psi^{\mathbf{C}}_h$ is the $c$-field for the hot reservoir $(h)$, which is at a temperature $T_h$. The initial state of the hot reservoir is prepared exactly in the same way as the working fluid, using the SPGPE, Eq.~(\ref{eq:SPGPE}), except with the subscript $s$ replaced by $h$.

\begin{figure}[t!]
    \centering
    \includegraphics[scale=0.7]{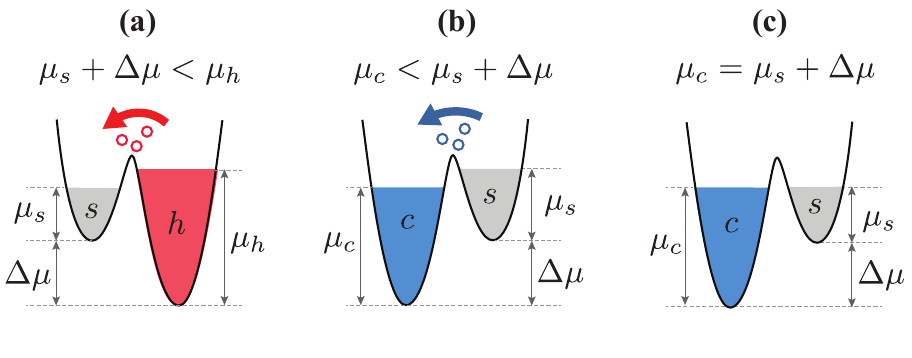}
    \caption{Illustration of three typical chemical potential offset ($\Delta \mu$) settings relevant for modelling the thermalization strokes when the working fluid is tunnel-coupled to a hot ($h$, red) or cold ($c$, blue) reservoir via a transverse double-well potential. In panel (a), the chemical potentials satisfy $\mu_s+\Delta \mu < \mu_h$, resulting in net particle flow from the reservoir to the working fluid, which is the case for stroke \textbf{AB}; panel (b) corresponds to the reverse situation, which is the case for stroke \textbf{CD} with $\mu_s+\Delta \mu > \mu_c$; finally, panel (c) illustrates a balanced situation of $\mu_s+\Delta \mu = \mu_c$, which does not result in any net particle flow between the working fluid and the cold reservoir (with the same being true if the cold reservoir is replaced by the hot one). }
    \label{fig:Chemical_offset}
\end{figure}

To initiate equilibration stroke \textbf{AB}, the tunnel-coupling parameter is quenched from $J=0$ to a constant value, which we chose to be $J=2 \omega$ for definiteness. Moreover, during the equilibration stroke with the hot reservoir, the interaction strength parameter of the working fluid is kept constant at value $g_s = g_h$. The external trapping potential in Eq.~(\ref{eqn:coupled_PGPEs_b})\,(b) is given by, 
\begin{equation}
\label{trap_chemicalshift}
    V_h(x) = \frac{1}{2} m \omega^2x^2 - \Delta \mu,
\end{equation}
where we have introduced a chemical potential offset, $\Delta \mu$, in order to control the net flow of particles from the reservoir to the working fluid if $\mu_s+\Delta \mu < \mu_h$, or vice versa -- from the working fluid to the reservoir -- if we were to chose $\mu_s+\Delta \mu > \mu_h$ (see Fig.~\ref{fig:Chemical_offset} for illustrations). The latter arrangement, with $\mu_s+\Delta \mu > \mu_c$, is the case for the stroke \textbf{CD} (see below).
We note here that, while the choice of $\Delta \mu$ affects the overall \emph{net} flow of particles upon completion of the \textbf{DA} stroke, transient transfer and oscillations of particles between the working fluid and the reservoir is still possible, and is in fact required for establishing eventual mutual thermal equilibrium.

The coupled PGPEs,~ Eqs.~(\ref{eqn:coupled_PGPEs_a}) and (\ref{eqn:coupled_PGPEs_b}), are evolved for a duration of time, which we refer to as thermalization time $t_\mathrm{th}$, during which the working fluid comes into thermal equilibrium with the hot reservoir. In the process, energy, $E_\mathrm{in} = \langle \hat{H}_s \rangle_{\textbf{B}}-\langle \hat{H}_s \rangle_{\textbf{A}}>0$, is transferred from the hot reservoir to the working fluid.

~

\noindent \textit{Step 3: Unitary expansion work stroke,} \textbf{BC}: After completion of the equilibration stroke with the hot reservoir, we switch off the tunnel coupling, i.e. set $J=0$, and perform the expansion work stroke by evolving the working fluid according to the PGPE given in Eq.~(\ref{eqn:PGPE}), except that now the interaction strength, $g(t)$ is decreased from the value $g_h$ back to $g_c$ according to the following linear protocol:
\begin{equation}\label{eq:g_exp}
    g(t) = g_{h}-(g_{h}-g_{c}) t  / t_\mathrm{w}.
\end{equation}
During the expansion work stroke, which is completed in the same duration ($t_\mathrm{w}$) as the compression stroke, the working fluid performs work $W_\mathrm{exp} < 0$. Similarly to $W_\mathrm{com}$, the quantity $W_\mathrm{exp}$ is computed numerically by evaluating the change in the Hamiltonian energy of the working fluid after completion of the expansion stroke at point \textbf{B}, i.e, $W_\mathrm{exp}=\langle \hat{H}_s \rangle_{\textbf{B}}-\langle \hat{H}_s \rangle_{\textbf{A}}$.

~

\noindent \textit{Step 4: Equilibration with cold reservoir,} \textbf{CD}: On completion of the expansion stroke, we enable tunnel coupling with the cold reservoir and simulate the dynamics of equilibration using the coupled PGPEs given in Eqs.~(\ref{eqn:coupled_PGPEs_a}) and (\ref{eqn:coupled_PGPEs_b}), except that the subscript $h$ is replaced by $c$. During this stroke, energy $E_\mathrm{out} = \langle \hat{H}_s \rangle_{\textbf{D}}-\langle \hat{H}_s \rangle_{\textbf{C}}<0$, is being transferred from the working fluid to the cold reservoir, returning the working fluid and the overall Otto cycle to the point of initialization \textbf{D}.
  
~ 

The overall performance of this engine cycle may be quantified through the net work, 
\begin{equation}
W =  W_\mathrm{{com}} + W_\mathrm{{exp}}, 
\end{equation}
which must be negative for the Otto cycle to operate as an engine, efficiency, 
\begin{equation}
\eta = - W/E_\mathrm{in}, 
\end{equation}
and power output, 
\begin{equation}
P = -W/t_\mathrm{tot}, 
\end{equation}
where $t_\mathrm{tot}$ is the total cycle time, given by $t_\mathrm{tot} = 2 (t_\mathrm{w} + t_\mathrm{th})$.

\section{Quantifying the timescales for the unitary work strokes} \label{sec:3}

\begin{figure}[t!]
    \centering
    \includegraphics[scale=0.38]{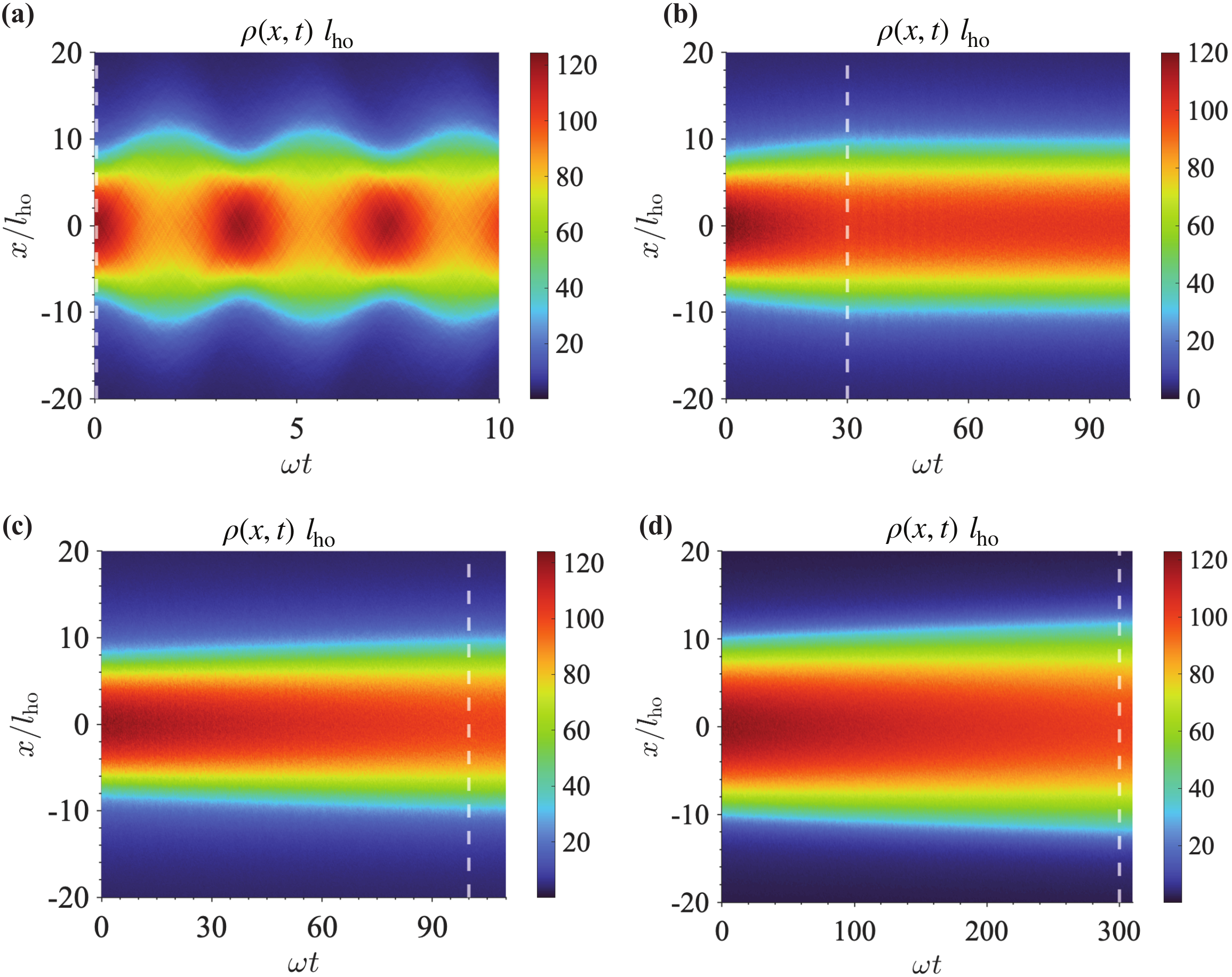}
    \caption{Time-evolution of the real-space density profile $\rho(x,t)$ of a weakly interacting 1D Bose gas, following an interaction strength quench. After preparing the initial state $(t = 0)$, a linear quench of the interaction strength $g_s$ was performed according to Eq.~(\ref{eq:g_com}), from an initial value of $g_c$ to the final value $g_h=1.80g_c$, and the working fluid was evolved according to the 1D PGPE (\ref{eqn:PGPE}). 
    The quench duration (shown as vertical dashed lines) was chosen as: (a) $t_\mathrm{w} = 0.05/\omega$ (sudden quench); (b) $t_\mathrm{w} = 30.0/\omega$ (intermediate); (c) $t_\mathrm{w} = 100/\omega$ (intermediate); and (d) $t_\mathrm{w} = 300/\omega$ (near-adiabatic or quasistatic quench). The dimensionless position $x/l_{\mathrm{ho}}$ is defined using the characteristic harmonic oscillator length along the longitudinal dimension, $l_{\mathrm{ho}} = \sqrt{\hbar/m \omega}$. All results in this paper are for a gas of $^{87}$Rb atoms with mass $m \simeq 1.44 \times 10^{-25}$ kg and 3D scattering length $a_\mathrm{s} \simeq 5.31\,$nm. The initial thermal equilibrium temperature of the working fluid here is $T_s = 86.3$ nK and the total number of particles is $N_s \simeq 1750$ which was obtained using the chemical potential $\mu_s = 6.62\times10^{-31}$ J. The other relevant physical parameters that were used are as follows: $\omega/2\pi = 20.0$ Hz and $g_c = 1.27 \times 10^{-38}$ J$\cdot$m. The transverse confinement frequency that was used to achieve this value of interaction strength was $\omega_{\perp}/2\pi = 1.81$ kHz (recall that $ g_s \!\simeq\! 2 \hbar \omega_{\perp}a$).} 
     \label{fig:Fig3_Turbo}
\end{figure}

In this section, we analyze the three distinct time-scales over which the work strokes may be performed via a quench of the interaction strength. These include: (\emph{i}) the sudden quench, where the efficiency is lowest due to the production of maximum irreversible work \cite{abah2019shortcut, rezek2006irreversible}; (\emph{ii}) the near-adiabatic (quasistatic) quench, which corresponds to near-maximal efficiency but minimal power output due to extremely long driving time \cite{born1928beweis}; and (\emph{iii}) the intermediate quench, which lies in between the first two extremes and highlights the trade-off between power and efficiency.

In theory, a sudden quench is often treated as if it were instantaneous \cite{bayocboc2023frequency, bouchoule2016finite, thomas2021thermalization}. However, in practice, a ``sudden'' quench is not truly instantaneous but occurs over a finite duration \cite{schemmer2018monitoring}. This duration must be fast enough to be nearly sudden with respect to the characteristic timescale for longitudinal dynamics, $t_{\parallel}=2\pi/\omega$, yet slow enough (near adiabatic) in relation to the characteristic timescale for transverse dynamics, $t_{\perp}=2\pi/\omega_{\perp}$, so that one avoids exciting any transverse excitations and maintains the 1D nature of the system \cite{schemmer2018monitoring}.

Accordingly, in our simulations, the unitary work strokes in the sudden quench regime are completed in a finite time by performing a linear quench of the interaction strength described in Eqs.~(\ref{eq:g_com}) and (\ref{eq:g_exp}), satisfying the following the condition:
\begin{equation}
\label{eqn:t_sudden}
  t_{\perp} \ll  t_\mathrm{w} \ll t_{\parallel}.
\end{equation}
The quasistatic or near-adiabatic engine cycle, on the other hand, corresponds to completing the work strokes over timescales during which the system remains approximately in thermal equilibrium states. This implies that the work strokes are near-adiabatic relative to both time-scales introduced above, i.e.,
\begin{equation}
\label{eq:t_ad}
     t_{\perp} \ll  t_{\parallel} \ll t_{\mathrm{w}}.
\end{equation}
Finally, to complete the work strokes in the intermediate regime, we follow the following condition:
\begin{equation}
\label{eq:t_inter}
   t_{\perp} \ll t_{\parallel} \sim t_\mathrm{w}.
\end{equation}

In Fig.~\ref{fig:Fig3_Turbo}, we demonstrate the time-evolution of the real-space density profile of the working fluid as we quench the interaction strength to perform the unitary \emph{compression} work stroke \textbf{CD} in time $t_\mathrm{w}$. (The dynamics during the \emph{expansion} strokes are similar and will not be shown). To identify distinct dynamical regimes as a function of the quench duration, we complete the interaction strength quench over various values of $t_\mathrm{w}$ and then continue simulating the unitary post-quench dynamics of the system (instead of immediately implementing the next stroke of the Otto cycle -- the thermalization stroke \textbf{BA}). This is done for diagnostics purposes only -- as to identify the presence, or lack thereof, of any longitudinal excitations in the system -- for the quench to be regarded as sudden or intermediate, as opposed to near-adiabatic (quasistatic) quench.

The vertical dashed lines in Figs.~\ref{fig:Fig3_Turbo}\,(a)--(d) mark the completion of the interaction quench over duration $t_{\mathrm{w}}$. In Fig.~\ref{fig:Fig3_Turbo}\,(a), this duration is too short for the dotted line to be visible; here, we are in the sudden quench regime and can clearly see excitation of breathing mode oscillations \cite{ fang2014quench, tschischik2013breathing,bayocboc2023frequency, schemmer2018monitoring, schmitz2013quantum,bouchoule2016finite} after the interaction quench is completed. 
In contrast, Fig.~\ref{fig:Fig3_Turbo}\,(b) and (c) depict the dynamics after the intermediate quench. Here, post-quench breathing mode oscillations are significantly suppressed, however, we can still observe weak breathing mode excitations during the quench, indicating that these quenches are not yet slow enough to be classified as near-adiabatic or quasistatic.
Finally, Fig.~\ref{fig:Fig3_Turbo} (d) illustrates what we classify as a near-adiabatic (quasistatic) quench. Over these timescales, no observable non-adiabatic excitations are produced. This regime of the Otto engine is expected to result in near-maximum efficiency due to minimal irreversible work produced during the work strokes \cite{rezek2006irreversible, abah2019shortcut, ccakmak2016irreversibility, keller2020feshbach}.

\section{Operational timescales for thermalization strokes under various dynamical scenarios} \label{sec:4}
The power output of an engine cycle is inversely dependent on the total cycle time. This is broken down into a sum of the duration of the unitary work strokes, which are controlled externally, and the equilibration strokes, which are not externally controlled but are typically assumed to be fast in comparison to that of the work strokes \cite{keller2020feshbach, boubakour2023interaction, li2018efficient, li2021shortcut, ccakmak2019spin, rezek2006irreversible, kosloff2017quantum}. In this section, we analyze these equilibration strokes in order to determine the characteristic operational timescale that governs the thermalization of the working fluid. In particular, we explore the effects of various factors, such as the size and temperature of the reservoir, as well as the duration of the prior work stroke which can leave the system in a highly non-equilibrium state after a sudden quench; or in a near-equilibrium state after a quasistatic quench.

\begin{figure}[tp]
    \centering
    \includegraphics[scale=0.38]{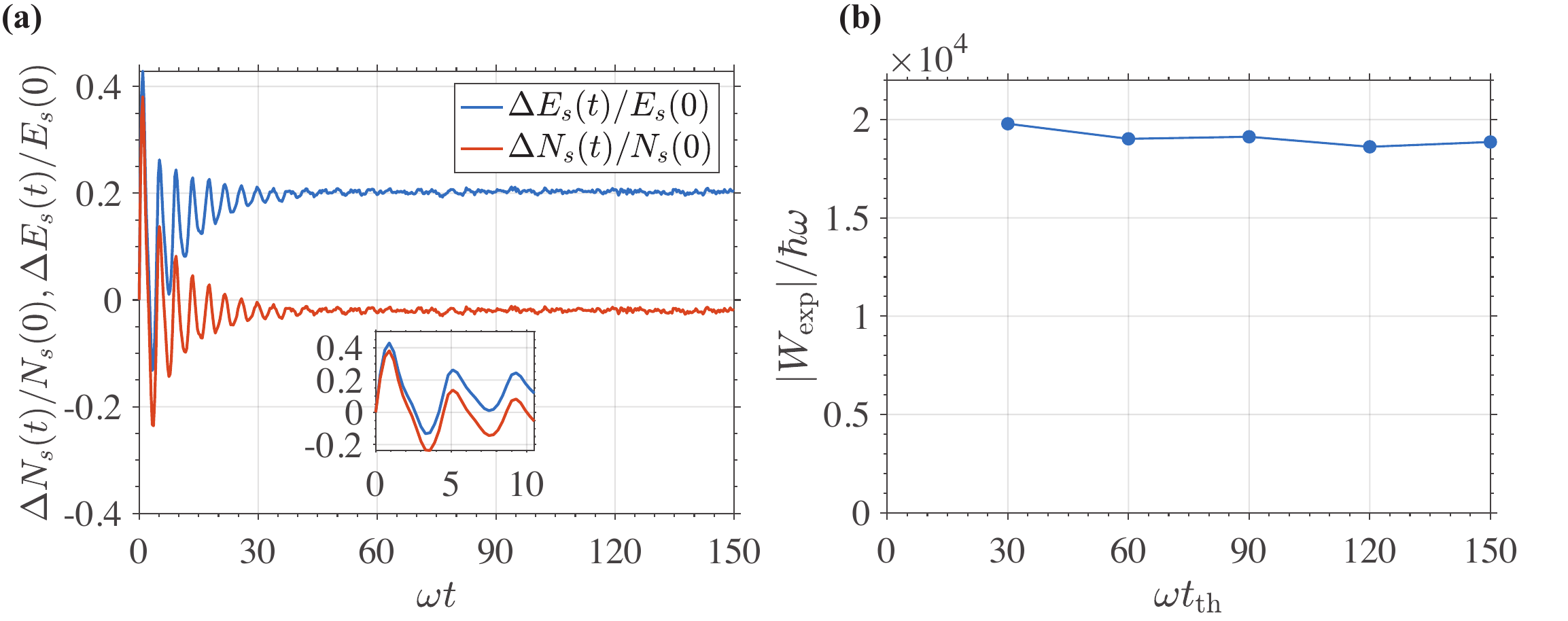}
    \caption{Relative change in energy and particle number, $\Delta E_s(t)/E_s(0)$ and $\Delta N_s(t)/N_s(0)$, of the working fluid versus dimensionless time $\omega t$ during the thermalization stroke \textbf{AB}. (b) Absolute value of the work extracted from the working fluid in the sudden-quench expansion stroke \textbf{BC} as a function of the dimensionless duration of the thermalization stroke \textbf{AB} $\omega t_{th}$. The dimensionless duration of expansion stroke \textbf{BC} is $\omega t_\mathrm{w}=0.05$. All parameter values for the initial state of the system and the ratio of the values of $g_c$ and $g_h$ are the same as in Fig.~\ref{fig:Fig3_Turbo}, except that now (during the expansion stroke) the interaction strength is quenched from $g_h$ back to $g_c$. The hot reservoir is prepared with interaction strength $g_h = 2.29 \times 10^{-38}$ J$\cdot$m and temperature $T_h=258$ nK. The ratio of particle number between the reservoir and the system is $N_h/N_s \simeq 7.31$. The particle number of the reservoir is $N_h \simeq 12800$, which is obtained using the chemical potential $\mu_h=39.7 \times 10^{-31}$ J. The ratio of temperature between the initial states of the system and the hot reservoir is $T_h/T_c =3.00$. The results in this figure correspond to the chemical offset arrangement shown in Fig.~\ref{fig:Chemical_offset}(c), where we have maintained $\Delta N_s (t \gg 1/\omega) \simeq 0$, by using $\mu_h \simeq \mu_s + \Delta \mu$, with $\Delta \mu = 29.5  \times 10^{-31}$ J.} 
   
    \label{fig:Fig3}
\end{figure}

\begin{figure}[tp]
    \centering
    \includegraphics[scale=0.39]{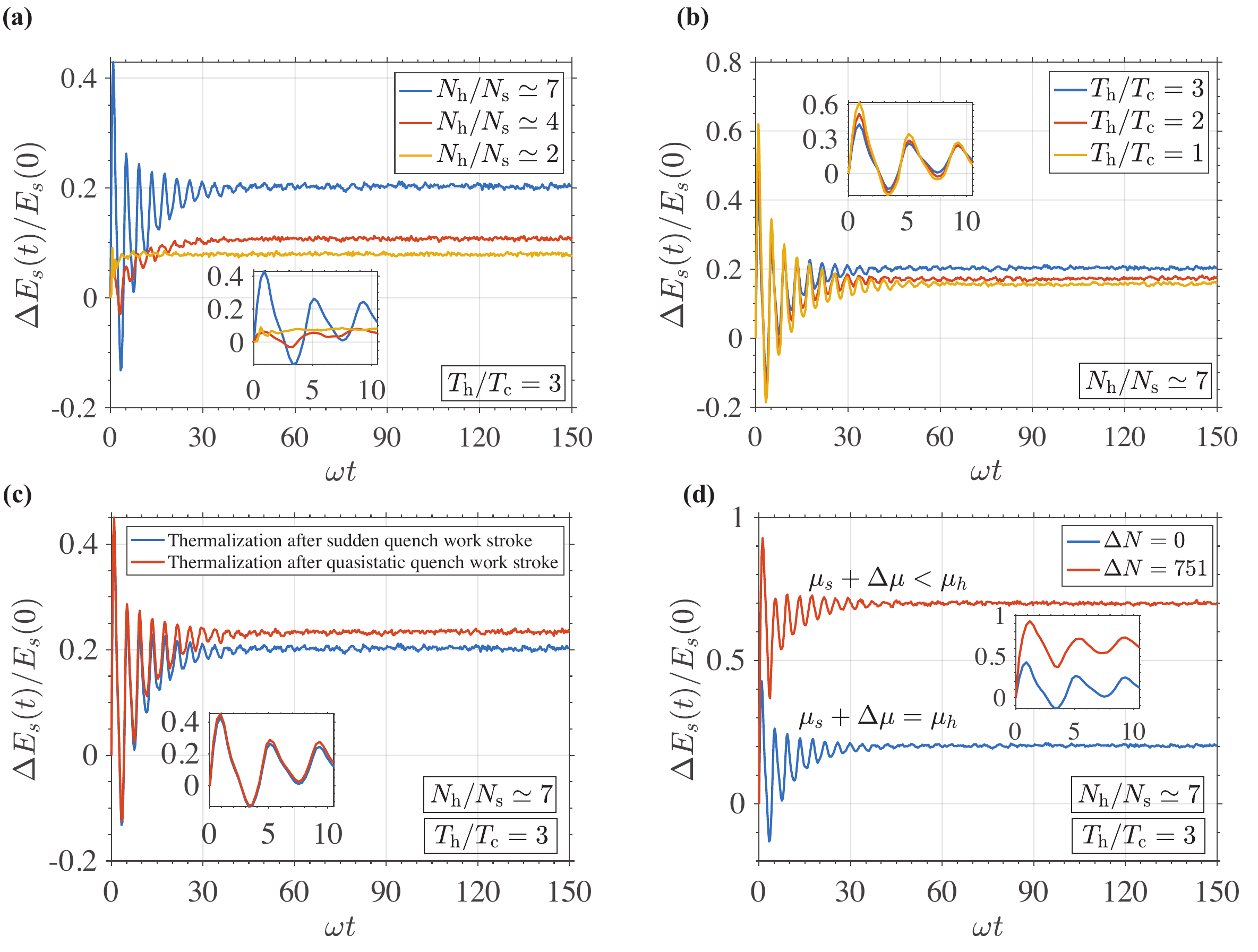}
    \caption{Relative change in energy of the working fluid, $\Delta E_s(t)/E_s(0)$, versus the dimensionless time $\omega t$ during the thermalization stroke \textbf{AB}, for various scenarios: (a) different ratios of particle number, $N_h/N_s \simeq 7$, $N_h/N_s \simeq 4$, and $N_h/N_s \simeq 2$; (b) different temperature ratios, $T_h/T_c = 3$, $T_h/T_c = 2$, and $T_h/T_c = 1$; (c) dynamics of thermalization following a sudden quench in the work stroke ($t_{\mathrm{w}} = 0.05/\omega$) compared to a quasistatic quench ($t_{\mathrm{w}} = 300/\omega$); and (d) dynamics of thermalization in a pure heat engine scenario ($\Delta N = 0$) compared to a chemical engine scenario with a net particle flow of $\Delta N = 751$ into the system from the hot reservoir. In all cases, the initial parameters of the system and the ratio of interaction strength quench for the work stroke are the same as in Fig.~\ref{fig:Fig3_Turbo}. In all panels, the system's initial particle number, $N_s\simeq 1750$ and temperature, $T_s=T_c=86.3$ nK were the same. For the remaining parameter values, see \cite{Parameter_values}.
      }
    \label{fig:Fig4}
\end{figure}

Specifically, we focus on the thermalization stroke \textbf{AB} that follows immediately after the unitary compression stroke \textbf{DA} studied in the previous section.
In Fig.~\ref{fig:Fig3}\,(a), we illustrate the time evolution of relative changes in the working fluid's (or system's, for which we use the subscript $s$) energy and particle number, $\Delta E_s(t)/E_s(0) = (E_s(t) - E_s(0))/E_s(0)$ and $\Delta N_s(t)/N_s(0) = (N_s(t) - N_s(0)/N_s(0)$, during this thermalization stroke. Here, the quantities $E_s(t) = \langle \hat{H_s}(t)\rangle$ and $N_s(t) = \int \langle \hat{\Psi}^{\dagger}_s (x,t)\hat{\Psi}_s (x,t) \rangle  dx$
are evaluated by replacing the creation and annihilation field operators by time-evolving stochastic realizations of the $c$-fields, $\Psi^*_{\textbf{C}}(x,t)$ and $\Psi_{\textbf{C}}(x,t)$, and by replacing quantum mechanical ensemble averages by stochastic averages over a large number (typically 2000) of stochastic realisations.
In the examples shown in Fig.~\ref{fig:Fig3}\,(a), we consider the scenario where the thermalization stroke \textbf{AB} is initiated immediately after executing the work stroke \textbf{DA} via a finite-time sudden quench (as in Fig.~\ref{fig:Fig3_Turbo}(a)); completion of the work stroke at $t=t_\mathrm{w}$, corresponds to the start of the thermalization stroke \textbf{AB}, which we reset to be the zero, $t= 0$, in the horizontal axes labels. 
 We observe that, after an initial rapid exchange of particles, which occurs in the form of damped oscillations, we have maintained---in this example---zero net exchange in the particle number of the working fluid, i.e. $\Delta N_{s}(t\gg 1/\omega) \simeq0$. This is achieved by tailoring the chemical potential offset, $\Delta \mu$, such that we maintain zero net flow of particles, as was illustrated in Fig.~\ref{fig:Chemical_offset}\,(c).
Even though $\Delta N_{s}(t\gg 1/\omega) \simeq0$, we observe a net increase in the energy of the working fluid after a sufficiently long duration of the thermalization stroke, $\Delta E_\mathrm{s}(t\geq 40/\omega)$, which is due to purely temperature imbalance. We point out here that, even though the working fluid may have not come to complete thermal equilibrium at time $t\sim 40/\omega$, the bulk of the energy transfer, which is to be converted into useful work in the next stroke \textbf{BC}, has already taken place by this time.

In Fig.~\ref{fig:Fig3}~(b), we show how the duration of the thermalization stroke \textbf{AB} affects the amount of work, $W_\mathrm{exp} <0$, done by the working fluid during the subsequent expansion stroke \textbf{BC}. More specifically, we show the absolute value of the work, $|W_\mathrm{exp}|$, as a function of the duration, $t_\mathrm{th}$, of the thermalization stroke \textbf{AB} during which the working fluid is kept in contact with the hot reservoir. We observe that after $t_\mathrm{th} \simeq 40/\omega$ the absolute work $|W_\mathrm{exp}|$ shows negligible further change with an increased duration of the thermalization stroke \textbf{AB}. This is consistent with the observation made in Fig.~\ref{fig:Fig3}\,(a) that the bulk of the energy from the hot reservoir, that is being converted into work, has already been transferred to the working fluid by $t_\mathrm{th}\simeq 40/\omega$.

In Figs.~\ref{fig:Fig4}\,(a) and (b) we illustrate, respectively, the effects of the size and temperature of the hot reservoir on the dynamics of the relative change in energy, $\Delta E_s (t)/E_s(0)$, of the working fluid during the thermalization stroke \textbf{AB}. In both these cases, just as we begin the thermalization stroke, the working fluid is in an out-of-equilibrium state after the completion of the compression work stroke via a sudden quench ($t_\mathrm{w}=0.05/\omega$). The completion of the work stroke at $t=t_\mathrm{w}$, corresponds to the start of the thermalization stroke \textbf{AB}, which we reset to be the zero, $t= 0$, in Fig.~\ref{fig:Fig4}, just as we did in Fig.~\ref{fig:Fig3}.

In the examples of Fig.~\ref{fig:Fig4}(a), the thermalization stroke is implemented using three different hot reservoirs, each with a different particle number but an identical temperature, whereas the number of particles in the working fluid is kept constant at $N_s=1750$. The energy transfer from the reservoir to the working fluid again takes place through damped oscillations.
The net increase in energy of the working fluid after a sufficiently long duration of the thermalization stroke, $\Delta E_s(t\!\gg1/\!\omega)$, increases with the size of the hot reservoir, as expected. Furthermore, when using a small reservoir, the amplitude of oscillations, responsible for energy transfer, is smaller than in the case of a larger reservoir, and the oscillations damp faster. 
This means that, when using a smaller reservoir, the thermalization stroke can be terminated at an earlier time, since the bulk of the energy transfer has already occurred during the first few oscillations. 

For the examples of Fig~\ref{fig:Fig4}~(b), we employ the same procedure as in Fig.\ref{fig:Fig4}(a), but in this case we consider three different temperatures of the hot reservoir, and keep its total particle number fixed. In these three scenarios, we observe that the dynamics of the relative change in energy of the working fluid are largely insensitive to the temperature of the reservoir and that the bulk of energy transfer from the reservoir to the working fluid occurs on the same timescale. This suggests that the temperature of the hot reservoir may not have a significant effect on the Otto engine power output. Additionally, we see that the net energy transfer from reservoirs of different temperatures does not vary much, suggesting that these temperatures may not have a significant effect on the efficiency of the engine either. This is consistent with a similar recent finding from Ref.~\cite{estrada2024quantum}, for a study of the conventional (volumetric) Otto cycle with partially condensed, harmonically trapped Bose-Einstein condensate as the working fluid.

In Fig.~\ref{fig:Fig4}\,(c), we analyse the dynamics of energy transfer from the reservoir to the working fluid, depending on whether the working fluid, after the previous work stroke, was left in a highly non-equilibrium state (such as in a sudden quench work stroke) or in a near-equilibrium state (such as after a quasistatic work stroke). Here, the curve corresponding to energy transfer after a sudden quench work stroke is the same as the respective curve from Fig.~\ref{fig:Fig4}\,(a). The curve corresponding to energy transfer after a quasistatic work stroke, on the other hand, is shown in red. As we see, the dynamics of energy transfer in both cases are very similar: thermalization in both cases occurs over similar time scales, and the net increase in energy of the working fluid is approximately the same, with the net energy increase after a quasistatic work stroke being only marginally larger than that after a sudden quench work stroke. This suggests that the rate and magnitude of energy transfer during the thermalization stroke does not strongly depend on the state (highly non-equilibrium versus near equilibrium) of the working fluid. Therefore, executing the work strokes in a quasistatic manner does not lead to faster thermalization nor a more significant increase in energy of the system at the end of the thermalization stroke with the hot reservoir.

Finally, in Fig.~\ref{fig:Fig4}~(d), we demonstrate our model's capability to function as a \textit{chemical} engine. 
We compare the relative increase in energy of the working fluid during the thermalization stroke \textbf{AB} in two scenarios: first, corresponding to a pure \emph{heat} engine, when there is no net flow of particles from the hot reservoir to the working fluid, i.e. $\Delta N_{s}(t\gg 1/\omega) \simeq0$, using the scheme of Fig.~\ref{fig:Chemical_offset}~(c) and repeating the result for $N_{h}/N_s=7$ from Fig.~\ref{fig:Fig3}\,(a) (blue curve); and second, corresponding to a \emph{chemical} engine, where we allow for an additional flow of $\Delta N$ particles from the hot reservoir into the working fluid, using the scheme of Fig.~\ref{fig:Chemical_offset}~(a) (red curve). We observe a significant increase in the energy of the working fluid as we perform additional chemical work via the particle inflow. Thus, increasing the particle inflow from the hot reservoir to the working fluid provides an opportunity to increase the beneficial net work, as we have more energy, $E_\mathrm{in}$, at our disposal to be utilized to perform work, $W_\mathrm{exp}$, in the subsequent expansion stroke.

\section{Performance of the full Otto cycle} \label{sec:5}

In this section, we combine the analysis of characteristic timescales for the work and thermalization strokes in the interaction-driven Otto cycle to evaluate the overall performance of the proposed Otto engine and the trade-off between its power and efficiency.

\subsection{Impossibility of operating as a heat engine}

\begin{figure}[tp]
    \centering
    \includegraphics[scale=0.4]{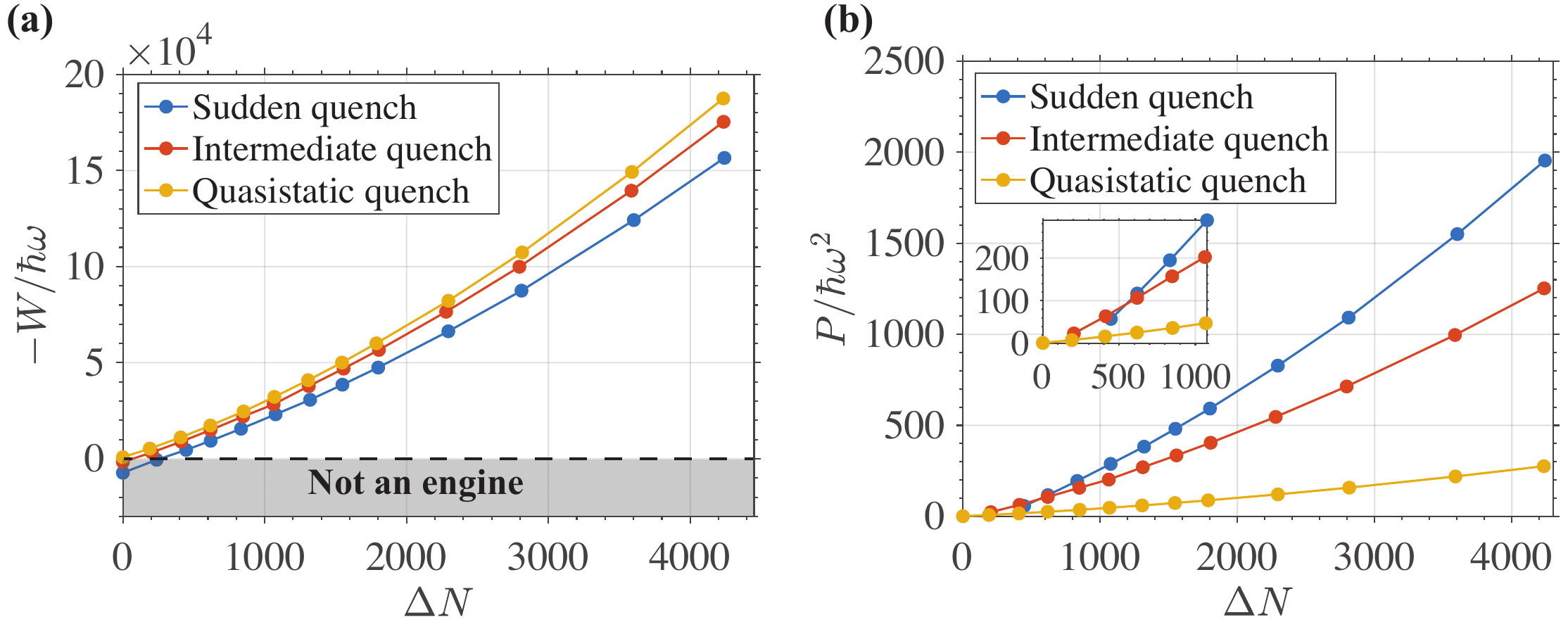}
    \caption{Net work and power output of an interaction-driven Otto engine as a function of the number of particles, $\Delta N$, exchanged with the hot and cold reservoirs during the thermalization strokes for sudden, intermediate, and quasistatic quenches. In panel (a), we show the net work $-W/\hbar \omega$ (in units of $\hbar \omega$), with $-W>0$ corresponding to positive net work done by the working fluid. For the first two data points of the blue curve and the first data point of the red curve, we have $-W<0$, meaning these points do not correspond to performance as an engine. In panel (b), we show the power output, $P/\hbar \omega^2$ (in units of $\hbar\omega^2$) of the Otto engine cycle corresponding to the data in (a). The data points that do not correspond to engine operation have been removed from (b). The duration of the thermalization stroke with each of the reservoirs was fixed at $t_\mathrm{th} = 40/\omega$ in all three cases. All other parameters are the same as in Fig.~\ref{fig:Fig3}. }
    \label{fig:Fig5}
\end{figure}

We first discuss a full interaction-quench Otto cycle with a harmonically trapped 1D Bose gas operating in a pure \emph{heat} engine setup, i.e., when the chemical potential offset $\Delta \mu$ (see Fig.~\ref{fig:Chemical_offset}) is chosen in such a way that there is no net particle flow between the working fluid and the reservoirs. This study was motivated by an attempt to extend the \emph{uniform} 1D Bose gas results of Refs.~\cite{watson2024interaction,chen2019interaction} to a harmonically trapped system, which is easier to realise experimentally. What we found, however, was that a harmonically trapped 1D Bose gas, unlike the uniform system, did not result in engine operation regime with large and negative $W<0$: the net work was either positive in the sudden quench scenario (implying that the system gained energy as a result of the cycle, rather than lost energy to a useful work), or was negative, but very small, in the quasistatic quench. This finding is illustrated in Fig.~\ref{fig:Fig5} (a), where see that $-W$ as a function of the number of particles $\Delta N$ is negative (i.e., $W>0$) for $\Delta N=0$, in the sudden-quench Otto cycle. The quantity $-W$ becomes positive (i.e., $W<0$) only at some finite $\Delta N$, in which case we refer to the Otto cycle as a \emph{chemical} engine (see next subsection), rather than \emph{heat} engine.

The main reason hindering the operation of the Otto cycle as a heat engine using a harmonically trapped 1D Bose gas is that the net work is now not only a function of the difference of the local atom-atom correlation functions at the hot and cold thermal equilibrium points \textbf{B} and \textbf{D} of the Otto cycle diagram of Fig.~\ref{fig:Fig1} (for details, see \cite{watson2024interaction}), but it also depends on the inhomogeneity of the density profile and, in particular, on the peak density at these equilibrium points. The overall effect of this is that, while the atom-atom pair correlation has a favourable dependence on the temperature $T$ for the net work done by the fluid to be large and positive, the dependence of the peak density on the temperature is not favourable and it cancels out the positive net work that would be otherwise realisable in a uniform system where the densities in the hot and cold equilibrium points are the same.

\subsection{Operating as an Otto \textit{chemical} engine}

Although operating our interaction-driven Otto cycle as a purely heat engine is not feasible using a harmonically trapped Bose gas, we find that one can still operate it as a chemical engine by performing additional chemical work on the working fluid during the thermalization stroke \textbf{AB}. This additional chemical work is performed via the inflow of particles $\Delta N$ from the hot reservoir to the working fluid, using the chemical potential offset arrangement shown in Fig.~\ref{fig:Chemical_offset}~(a) and demonstrated in Fig.~\ref{fig:Fig4}~(d). The increase in total particle number results in a corresponding increase in the energy of the working fluid, which is available to be converted into mechanical work in the subsequent work stroke \textbf{BC}. After completing this work stroke, we couple the working fluid to the cold reservoir and transfer the same excess number of particles $\Delta N$ to the cold reservoir in the equilibration stroke \textbf{CD}, hence returning the working fluid into the state with the same initial number of particles.

The efficiency, $\eta = -W/E_\mathrm{in}$ of such a chemical Otto engine can be calculated by simply evaluating the energy differences at the end of each stroke, as described in Sec.~\ref{sec:2}. Unlike the case of a pure heat engine, the energy, $E_\mathrm{in} = \langle \hat{H}_s \rangle_{\textbf{B}}-\langle \hat{H}_s \rangle_{\textbf{A}}>0$, in an Otto chemical engine includes a contribution from chemical work and can be expressed as
\begin{equation}
    E_\mathrm{in} = Q_\mathrm{in} + W_\mathrm{chemical}.
\end{equation}
Here, $Q_{\mathrm{in}}$ is the heat taken in by the working during the thermalization stroke with the hot reservoir, \textbf{AB}, whereas $W_\mathrm{chemical}$ represents the additional chemical work done on the working fluid via the transfer of $\Delta N$ particles.
Though calculating the individual contributions, $Q_{\mathrm{in}}$ and $W_{\mathrm{chemical}}$, in the total $E_{\mathrm{in}}$ is a nontrivial task, as the heat and particle transport are intrinsically coupled processes (see, e.g., \cite{brantut2013thermoelectric, husmann2018breakdown}), we emphasise here that the chemical work is included in the overall energetic cost of evaluating the efficiency of the Otto chemical engine by defining the efficiency via 
$\eta = -W/E_\mathrm{in}$, rather than via $\eta = -W/Q_\mathrm{in}$.

\subsection{Trade-off between power and efficiency}

We finally analyse the overall performance of the finite-time Otto chemical engine and evaluate the trade-off between its power and efficiency.
In Fig.~\ref{fig:Fig5}, we show the net work and power output of the Otto chemical engine as a function of the number of particles $\Delta N$ exchanged with the reservoir, for three types of work strokes, corresponding to: sudden interaction quench, intermediate quench, and quasistatic quench. We see that, as we increase $\Delta N$, both the net work and power output increase in all three scenarios. Furthermore, as evident in Fig.~\ref{fig:Fig5}\,(a), the maximum work output occurs when the unitary work strokes are implemented through a quasistatic (near-
adiabatic) quench of the interaction strength, as expected. We observe, however, that transitioning from the slowest quasistatic quench to the fastest sudden quench regime results in a
relatively minor loss in work output, despite the work stroke being executed orders
of magnitude faster. This suggests that in the explored engine cycle, non-adiabatic
excitations contribute minimally to irreversible work and therefore optimising the quench protocol via a shortcut to adiabaticity \cite{del2013shortcuts,del2019focus} may not be necessary to operate at near-maximum efficiency \cite{li2018efficient, keller2020feshbach, abah2019shortcut, fogarty2020many}.

In Fig.~\ref{fig:Fig5}~(b), which shows the power, we observe that beyond a specific threshold value of $\Delta N \simeq 530$, the engine driven by a sudden quench of the interaction strength achieves a higher power output compared to the engine driven by a slow quasistatic and an intermediate quench for the work strokes. This increased power output can be attributed to the combination of two factors: first, the total engine driving time, $t_\mathrm{tot}$, in the sudden quench scenario is significantly shorter than the intermediate and the quasistatic quench; second, as demonstrated in Fig.~\ref{fig:Fig5}~(a), in a sudden quench engine, there is a relatively minor loss in net work $-W$, despite the work strokes being executed significantly faster. Consequently, given that power output is defined as $P = -W/t_\mathrm{tot}$, the sudden interaction quench engine significantly outperforms both the quasistatic and intermediate quench engines in terms of power output.

\begin{figure}[tbp]
    \centering
    \includegraphics[scale=0.40]{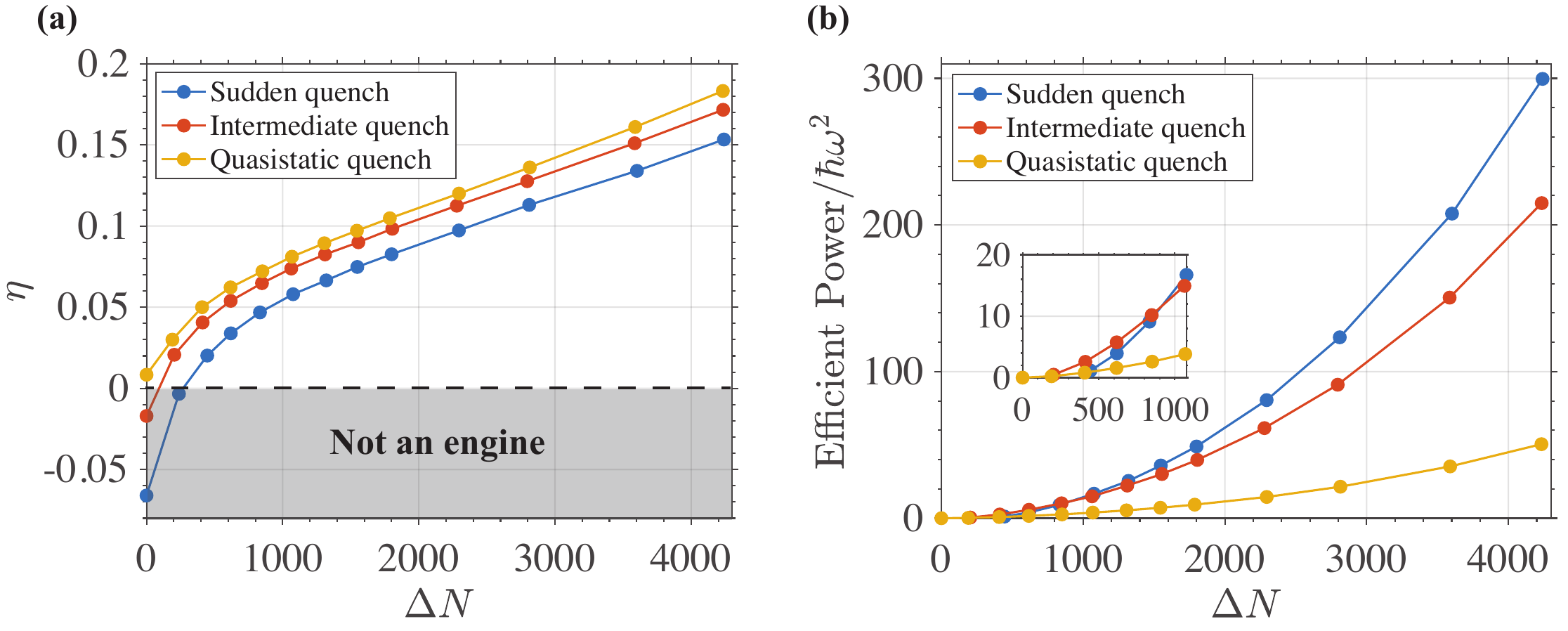}
    \caption{Efficiency, $\eta$, and efficient power as a function of the number of particles $\Delta N$ exchanged with the hot and cold reservoirs during the thermalization strokes for sudden, intermediate, and quasistatic quenches. All initial parameters of the working fluid and the reservoir are the same as in Fig.~\ref{fig:Fig3}. The thermalization time with the reservoir was fixed at $t_\mathrm{th} = 40/\omega$ for all three cases. In panel (a), we show the efficiency $\eta$, with $\eta>0$ corresponding to operation as an engine, i.e., when net work $-W >0$; the first two data points of the blue curve and the first data point of the red curve in (a) do not correspond to engine operation, as $-W < 0$ and hence $\eta<0$. In panel (b), we show the efficient power (see text) of the Otto engine cycle, corresponding to the data in (a). The data points that do not correspond to engine operation have been removed from (b). } 
    \label{fig:Fig6}
\end{figure}

In Fig.~\ref{fig:Fig6}~(a), we plot the efficiency as a function of the number of particles, $\Delta N$, for the chemical Otto engine driven by sudden, intermediate, and near-adiabatic (quasistatic) quenches of interaction strength during the work strokes. Consistent with observations in work and power output, an increase in $\Delta N$ leads to enhanced efficiency for all three quench scenarios. Furthermore, we observe that the sudden quench engine operates at efficiencies that are very close to the near-maximum limit achieved by the quasistatic work strokes. This result aligns with the results shown in Fig.~\ref{fig:Fig5}~(a), and means one can operate at near-maximum efficiencies by implementing the simplest finite-time and even sudden work strokes without relying on any optimisation protocol such as the STA to achieve similarly high efficiency \cite{del2013shortcuts, keller2020feshbach,abah2019shortcut, li2018efficient, fogarty2020many}. The results of Figs.~\ref{fig:Fig5}~(a) and \ref{fig:Fig6}~(b) confirm that irreversible work due to non-adiabatic excitations is not significant in this model, thereby making the use of STA practically redundant in our proposed engine cycle.

Furthermore, to quantify the trade-off between efficiency and power, we use the parameter ``efficient power'', which was first proposed in Ref.~\cite{yilmaz2006new}. Efficient power is simply a product of the efficiency and power output, and provides a direct relation between the increase in the power output per unit decrease in the efficiency \cite{yilmaz2006new,myers2020bosons}. In Fig.~\ref{fig:Fig6}~(b), we show the efficient power for the engines driven by sudden, intermediate and quasistatic interaction quenches for work strokes as a function of $\Delta N$. We observed that the engine operating in the sudden quench regime provides the maximum efficient power provided a certain threshold value of $\Delta N$ (equal to $\Delta N \simeq 900$ in this example) is crossed.

The main takeaway from the finite-time analysis of the proposed Otto chemical engine cycle is the favourable trade-off between efficiency and power output achieved by executing the work strokes through a sudden quench of the interaction strength. Furthermore, as we increase the number of particles, $\Delta N$, exchanged with the reservoirs, we observe a boost in engine performance across all three chosen quench times of the work strokes considered in this study: sudden quench, intermediate, and quasistatic quench. We additionally note that the qualitative conclusions presented above remain valid as long as we are within the weakly interacting regime of the 1D Bose gas, irrespective of the specific values of the parameters chosen.

\section{Conclusions} \label{sec:conclusion}

We simulated a finite-time quantum Otto cycle driven by a quench of atomic interactions of a 1D Bose gas in the weakly interacting quasicondensate regime. Our analysis included a simulation of both the unitary work strokes and the thermalization strokes for the proposed Otto cycle. To simulate the work strokes, we treated the working fluid as an isolated many-body quantum system undergoing unitary evolution, starting from a thermal initial state. 
The thermalization strokes, on the other hand, were simulated by treating the working fluid as an open many-body quantum system coupled to another many-body quantum system serving as the reservoir, both treated microscopically. We identified characteristic operational timescales for these thermalization strokes in experimentally realistic regimes.

The Otto engine's performance was evaluated in three different scenarios corresponding to three typical timescales for the execution of the work strokes: a sudden quench, an intermediate quench, and a slow quasistatic (near-adiabatic) quench. We first found that, contrary to a uniformly trapped Bose gas \cite{watson2024interaction}, a harmonically trapped system does not function as a heat engine. Nonetheless, we have also found that engine operation can be restored by enabling additional chemical work in the form of particle inflow from the hot reservoir to the working fluid. Thus, we have found that a harmonically trapped 1D Bose gas can operate effectively as a \textit{chemical} Otto engine. We have shown that such a chemical Otto engine, when operating in the sudden quench regime, achieves an efficiency that is quite close to the near-maximum limit obtained by implementing the work strokes in a quasistatic fashion. Thus, in our proposed engine cycle, it is possible to operate at near-maximum efficiency by executing the simplest finite time quench (linear quench) or even a sudden quench of the interaction strength, without relying on optimization protocols such as the STA. The primary reason for this is the minimal amount of irreversible work generated by non-adiabatic excitations during the finite-time driving of the Hamiltonian to execute the work strokes. Hence, when work strokes are executed through a sudden quench of the interaction strength, we observe a favourable trade-off between efficiency and power output in the \textit{chemical} Otto engine explored in this work.

In terms of future outlook, we note that the $c$-field approach employed here is limited to only the weakly interacting regime of the 1D Bose gas \cite{Blakie_cfield_2008}. Thus, future work could utilize alternate theoretical approaches, such as generalized hydrodynamics (GHD) \cite{castro2016emergent, bertini2016transport, watson2023benchmarks}, to explore the finite-time performance of the proposed engine across a much broader available parameter space of interaction strength and temperature in a 1D Bose gas \cite{kheruntsyan2003pair,kheruntsyan2005finite}. Further, conducting a similar analysis with a weakly interacting quasicondensate as the working fluid and a strongly interacting Tonks–Girardeau gas as the thermal reservoirs, or vice versa, could be another interesting scenario to address.

 This work was supported through Australian Research Council Discovery Project Grant No. DP190101515.

\section*{Acknowledgements}

 This work was supported through Australian Research Council Discovery Project Grant No. DP190101515.

\clearpage

\section*{References}

\bibliographystyle{unsrt}


\begin{thebibliography}{10}
\bibitem{Parameter_values}
In the three scenarios depicted in Fig.\ref{fig:Fig4}\,(a), the temperature of
  the hot reservoir, $T_h = 258$nK, was kept constant. The particle numbers
  were varied by using chemical potentials $\mu_h = 39.7 \times 10^{-31}$ J,
  $\mu_h = 26.5 \times 10^{-31}$ J, and $\mu_h = 17.2 \times 10^{-31}$ J,
  resulting in $N_h \simeq 12800$, $N_h \simeq 7420$, and $N_h \simeq 4010$,
  respectively. These cases correspond to the chemical offset configuration
  illustrated in Fig.~\ref{fig:Chemical_offset}(c), achieving $\Delta N_s (t
  \gg 1/\omega) \simeq 0$ by setting $\mu_h \simeq \mu_s + \Delta \mu$, where
  $\Delta \mu$ values are $29.5 \times 10^{-31}$ J, $17.4 \times 10^{-31}$ J,
  and $8.48 \times 10^{-31}$ J for each case, respectively. Similarly, in (b),
  the particle number $N_h \simeq 12800$ was kept constant while the
  temperature was varied with values of $T_h = 258$nK, $T_h = 172$nK, and $T_h
  = 86.3$nK. In the three cases shown in (b), $\Delta N_s (t \gg 1/\omega)
  \simeq 0$ was achieved by setting $\mu_h \simeq \mu_s + \Delta \mu$, where
  $\Delta \mu$ values are $29.5 \times 10^{-31}$ J, $30.3 \times 10^{-31}$ J,
  and $30.3 \times 10^{-31}$ J for each case, respectively. In (a), (b) and
  (d), the thermalization stroke was implemented following a sudden quench for
  work stroke ($t_\mathrm{w} = 0.05/\omega$), where the system was in a highly
  out-of-equilibrium state (See Fig~\ref{fig:Fig3_Turbo}(a)). In (d), a
  particle flow of $\Delta N =751$ from the hot reservoir to the system was
  achieved by using $\mu_h > \mu_s + \Delta \mu$ (see Fig.
  \ref{fig:Chemical_offset}(a)) with $\Delta \mu =25.1 \times 10^{-31}$ J. In
  (c) and (d), the initial particle number and temperature of the hot reservoir
  were $N_h \simeq 12800$ and $T_h = 258$ nK, respectively.
\end{thebibliography}

\end{document}